\documentstyle[prl,aps,preprint,epsfig]{revtex}
\tightenlines
\def\bea{\begin{eqnarray}}
\def\eea{\end{eqnarray}}
\begin{document}
\title{The Bose molecule in one dimension}
\author{Joseph P. Straley$^{1}$ and Eugene B. Kolomeisky$^{2}$}
\address{$^{1}$Department of Physics and Astronomy,
University of Kentucky, Lexington, Kentucky 40506-0055}
\address{$^{2}$Department of Physics,
University of Virginia, 382 McCormick Road, P. O. Box 400714, 
Charlottesville, Virginia 22904-4714}
\maketitle
\begin{abstract}
We give the Green function, 
momentum distribution, two-particle correlation function, and structure factor 
for the bound state of $N$ indistinguishable bosons with an attractive 
delta-function interaction in one dimension,
and an argument showing that this boson "molecule" has no excited 
states other than dissociation into separated pieces.
\end{abstract}

\section{The Boson Molecule}
Consider a system of $N$ identical bosons in one dimension, with an 
attractive delta-function interaction.  There is a bound state for all $N$, 
having the form of a well-localized "molecule"\cite{Lieb,McGuire,Thacker}.  
This result 
is interesting, since it is the unique example of an exactly solved localized 
many-body system.  A possible experimental realization of this system would be in
a $^{7}Li$ vapor \cite{Bradley} confined to an atomic trap which is so 
tight in two directions, that the system can be regarded as one-dimensional
with respect to the third direction.  Such traps are realistic prospects in 
the near future \cite{Jackson}, and high-aspect ratio cigar-shaped traps 
approximating quasi-one-dimensional systems are already available 
experimentally \cite{Andrews}. 

We seek the localized eigenfunctions $\Psi \{$x$_{i}\}$ of the Hamiltonian 
\begin {equation}
\label{Eq:Eq11}
H = -\sum^{N}_{i=1} {\hbar ^{2}\over 2m} {d^{2}\over dx_{i}^{2}} - W\sum^{}_{i<j} \delta 
(x_{i} - x_{j})
\end{equation}
wherein $m$ is the particle mass and $W$ is a positive parameter representing 
the strength of interparticle attraction.

The wavefunction and many derived quantities can  be given exactly and in 
closed form; however, it is useful to first consider the mean-field version of 
this theory.  The Gross-Pitaevskii (GP) approximation \cite{GP} assumes 
that $\Psi \{x_{i}\}$ can be
written as a simple product of wavefunctions 
\begin{equation}
\label{Eq:Eq1aa}
\Psi ( \{ x_{i} \} ) = \prod_{i} \psi(x_{i})
\end{equation}
where the 
$\psi (x)$ are eigenfunctions
for a single particle moving in the average potential of all the others
\begin {equation}
\label{Eq:Eq1a}
\epsilon  \psi (x) = - {\hbar ^{2}\over 2m} {d^{2}\psi (x)\over dx^{2}} + V(x) \psi 
(x)  .
\end{equation}
where $V(x) = -(N-1)W |\psi _{0}(x)|^{2}$.  Thus the ground-state wavefunction $\psi 
_{0}$ is
actually determined by solving a nonlinear Schr\"odinger equation.  
We will show that this approximation generally captures the physics
of the large $N$ limit; this is in contrast to the case of repulsive
bosons in one dimension, for which the interaction terms must be
treated more carefully \cite{KNSQ}. 
The
solution to Eq.(3) is
\begin {equation}
\label{Eq:Eq1b}
\psi _{0}(x) = \sqrt{{(N-1)\alpha \over 4}} {1\over \cosh((N-1)\alpha  x/2)}
\end{equation}
where $\alpha  = mW/\hbar ^{2}$; the single particle energy is $\epsilon  = - \alpha 
(N-1)^{2}W/8$.  This has the 
implications for the many-body problem that there can be a state that is 
localized in the sense that it is unlikely that any particle be found far from 
the center of mass, and provides an estimate for the particle density (with 
the center of mass fixed at the origin) $n(x) = N\psi (x)^{2}$; the corresponding
ground state energy is $E = -\alpha N(N-1)^{2}W/24$.

It can be shown that there are no other bound states of Eq. (3) for the 
potential V(x) constructed from the ground state wavefunction.  This suggests 
that we cannot construct approximate localized excited states 
of the boson molecule by
symmetrizing $\psi (x_{1}) \prod^{N}_{i=2}\psi _{0}(x_{i})$ (for
some appropriate choice of $\psi$).

The GP wavefunction is not translationally invariant, and thus violates momentum 
conservation.  Every translation of it is a solution to the same self-consistent 
equation with the same energy; then by forming the linear combination
$\Psi_{S} (\{x_{i} \} ) = \int \Psi (\{x_{i} - X \} ) dX $ we can restore
this symmetry.  
\section{The exact wavefunction }
For the case $N = 2$, the wavefunction can be found by changing to relative
coordinates, giving a problem equivalent to a single particle and a fixed
potential well.  The (unnormalized) ground state wavefunction is
\begin {equation}
\label{Eq:Eq2}
\Psi _{2}(x_{1},x_{2}) = \exp(-{\frac 1 2}\alpha |x_{1}-x_{2}|)
\end{equation}
where $\alpha  = mW/\hbar ^{2}$; the corresponding energy is $E_{2} = - \hbar 
^{2}\alpha ^{2}/4m = - mW^{2}/4\hbar ^{2}$.  For 
$x_{1} \neq  x_{2}$ the delta function vanishes, and the wavefunction is an 
eigenfunction
of the kinetic energy operator; for a bound state this is the product of
exponential functions with real argument.  The potential generates the
discontinuity in the derivative of $\Psi$ at the crossing point $x_{1} = x_{2}$.  
These two
properties hold for all $N$, so that we can write down the general
(unnormalized) wavefunction as \cite{Thacker}
\begin {equation}
\label{Eq:Eq3}
\Psi _{N}(x_{1}, \ldots   ,x_{N}) = \exp(-(\alpha /2) \sum^{}_{i<j}|x_{i}-x_{j}|)
\end{equation}
For the sector $x_{1} \le  x_{2}\le  \ldots  \le  x_{N}$ this can also be written in 
the form
\begin {equation}
\label{Eq:Eq4}
\Psi_{N} = \exp((1-N) \alpha x_{N} /2 + (3-N) \alpha x_{N-1} /2 
+ \ldots + (N-1) \alpha x_{1} /2 )
\end{equation}
In view of the exchange symmetry of bosons, this completely describes the 
wavefunction.  The ground state energy is proportional to the sums of the 
squares of the coefficients of $x_{i}$ in the argument of the exponential; 
explicitly, \cite{Thacker}
\begin {equation}
\label{Eq:Eq5}
E_{N} = - (N^{3}-N)\alpha W/24 = - (N^{3}-N) mW^{2}/24\hbar ^{2}
\end{equation}
This agrees with the GP result only in leading order in $N$.
\section{The single particle density}
The implications of the wavefunction are more clearly revealed when reduced to 
the probability that there is a particle at position $x$ relative to the center 
of mass.  For the case $N=2$, there can only be a particle at $x$ if the other is
at $-x$, and then the probability is proportional to the square of the
wavefunction:
\begin {equation}
\label{Eq:Eq6}
P_{2} = 2\alpha  \exp(-2\alpha  |x|)
\end{equation}
This has been normalized so that the integral over all $x$ gives $N = 2$. 

For larger $N$ we integrate over the unconstrained degrees of freedom; for 
example for $N = 3$ the probability of finding a particle at $x > 0$ is proportional
to
\begin {equation}
\label{Eq:Eq7}
\int^{x}_{-x/2}\Psi ^{2}(-x-x_{2},x_{2},x) dx_{2} + \int^{\infty }_{x} \Psi ^{2}(-x-
x_{3},x,x_{3}) dx_{3}
\end{equation}
where the integrals describe the case that it is particle 3 or particle 2 that 
is at $x$ (particle 1 is necessarily in the region $x < 0$).  The resulting
normalized probability distribution is
\begin {equation}
\label{Eq:Eq8}
P_{3}(x) =3\cdot 2\alpha  [\exp(-3\alpha  |x|) - {1\over 2} \exp(-6\alpha  |x|)] = 
3\cdot 2\alpha [z - {\frac 1 2} z^{2}]
\end{equation}
where in the second representation we have introduced the abbreviation $z =
\exp(-N\alpha |x|)$.  
In a similar way $P_{N}$ can be constructed for other small $N$.   
The results for $N \le  7$ suggest the general expression (given previously by 
Yoon and Negele\cite{Yoon})
\begin {equation}
\label{Eq:Eq12}
P_{N}(x) = \alpha  \sum^{N-1}_{n=1} (-1)^{n+1} {n N!N!\over (N+n-1)!(N-
n-1)!} \exp(-nN\alpha  |x|)
\end{equation}
This has been normalized so that the integral over all $x$ gives $N$.
The density at $x = 0$ is $P_{N}(0) = N^{2}(N-1)/(4N-6)$.  An interesting feature of 
the
distributions is revealed when we represent them as a power series in $|x|$: the
coefficients of $|x|^{n}$ vanish when $n$ is odd and less than $2N-3$.  Thus the
discontinuity in slope at $x = 0$ that is present in $P_{2}$ is replaced by a
discontinuity in a much higher derivative for larger $N$ (see footnote 7).  This
property is a demonstrable consequence of the representation (12); in fact,
combined with the assumption that $P_{N}$ is a polynomial in $z$ 
of order $(N-1)$, this
is sufficient to determine the form of $P_{N}$ up to normalization.

In the limit of large $N$, Eq. (12) reduces to
\begin {equation}
\label{Eq:Eq13}
P_{N}(x) \approx  N^{2}\alpha  \sum^{N-1}_{n=1} (-1)^{n+1} n \exp(-nN\alpha  |x|) 
\approx  {N^{2}\alpha \over 4}$ sech$^{2}(N\alpha x/2)
\end{equation}
which has form similar to the GP result 
$P_{N}(x) = {(N-1)^{2}\alpha \over 4}$ sech$^{2}((N-1)\alpha x/2).$
Figure 1 shows $(1/N^{2}\alpha )
P_{N}(x)$ as a function of $N \alpha x$, for various $N$.

We may also characterize $P_{N}(x)$ through its moments.  Starting from (12) it can 
be shown that
\begin {equation}
\label{Eq:Eq14}
<x^{2}> = {1\over N} \int^{\infty }_{-\infty }x^{2} P_{N}(x) dx = \sum^{N-1}_{M=1} 
{2\over \alpha ^{2}N^{2}M^{2}} \approx  {\pi ^{2}\over 3\alpha ^{2}N^{2}}
\end{equation}
\begin {equation}
<x^{4}> = \sum^{N-1}_{M=1} \sum^{N-1}_{P=1}
{12\over \alpha^{4}N^{4}M^{2}P^{2}}
+ \sum^{N-1}_{M=1}
{12\over \alpha^{4}N^{4}M^{4}} \approx 
{7\pi ^{4}\over 15\alpha^{4}N^{4}}
\end{equation}
The GP theory would give $<x^{2}> ={\pi ^{2}\over 3\alpha ^{2}(N-1)^{2}}$ 
(and similarly for $<x^{4}>$) -- they agree at large $N$, but the exact result 
approaches the asymptotic limit from below, while GP approaches from above.
\section{Green function and momentum distribution}
The single-particle Green function $G_{N}$ is calculated from the wavefunction by
constructing the product $\Psi (x,x_{2},x_{3},..x_{N})\Psi ^{*}(0,x_{2},x_{3},\ldots  
x_{N})$ and integrating out the
spectator coordinates $x_{2},x_{3},\ldots  ,x_{N}$.   
Explicit construction of the $G_{N}$ for $N \le  6$ suggests the result
\begin{equation}
\label{Eq14a}
G_{N}(x) = {g_{N} \over 2N} \sum^{N}_{m=1} \{
2 + \alpha  |x| [N^{2}-1 - 4K(m)] \} z^{K(m)  }
\end{equation}
where $w = \exp(-\alpha |x|)$ and $K(m) = [N^{2}-1-(N+1-2m)^{2}]/4$.   
With the choice $g_{N} = 1$, $G_{N}(x)$ is normalized
so that $G_{N}(0) =
1$; a physically more meaningful normalization would be to choose 
$\int^{- \infty}_{\infty} G_{N}(x) dx = 1$, but we have
not succeeded in finding the closed form representation for the corresponding
$g_{N}$.
Since $K(N+1-m)
= K(m)$, the sum
contains every term twice, except for $m =(N+1)/2$ when $N$ is odd.

In the large $N$ limit, $K(m) \approx (N-1)(2m-1)/2 ,$ and then
\begin{equation}
\label{Eq14b}
G_{N} \approx N \alpha x / 2 \sinh ( N\alpha x /2) 
={\frac 1 2 } \int^{\infty }_{-\infty } \psi_{0} (x+y) \psi_{0} (y) dy
\end{equation}
which is the Green function that one would construct from the
translationally invariant GP wavefunction $\Psi_{S}$.

The momentum distribution is the Fourier transform of G$_{N}$
\begin{equation}
\label{Eq:Eq14b}
G_{N}(k) = {1 \over 2N} \sum^{N}_{m=1} \{
2 {\alpha K\over \alpha ^{2}K^{2}+k^{2}} + {\alpha ^{2}K^{2}-k^{2}\over
[\alpha ^{2}K^{2}+k^{2}]^{2}} [N^{2}-1 - 4K]  \alpha \}
\end{equation}
where $k$ is the wavevector.
Figure 2 gives a graph of $N \alpha G_{N}(k)$ versus $k / N \alpha$,
for various $N$.
As in the case of the repulsive boson system\cite{Lieb},
the momentum distribution is smooth: there is no Bose condensation.
The width of the momentum distribution is set by the parameter
$\alpha$, so that the momentum distribution for the noninteracting case ($\alpha = 0$) is 
attained in a natural way.
In a highly localized wavefunction we would hardly expect to find
{\it long-range} off-diagonal order!

\section{Two-point correlation function}
The two-point correlation function is calculated from the squared wavefunction 
by choosing one of the particles to be $x_{i} = 0$ and another to be at $x_{j} = 
x$
(in all possible ways), and integrating out all other $\{ x_{i} \}$.  We have
chosen to normalize $C_{N}(x)$ so that the integral over all $x$ gives unity.  It
proved harder to analyze, because its representation is more irregular: for
example,
\begin {equation}
\label{Eq:Eq15}
C_{6}(x) = {\alpha \over 30 } [-110 w^{5} + 64 w^{8} + 81 w^{9} + \alpha  |x| (400 
w^{5} + 256 w^{8})]
\end{equation}
where again $w = \exp(-\alpha |x|)$.  
Based on the explicit construction of $C_{N}$ for $N \le  7$, we
find
\begin {equation}
\label{Eq:Eq16}
C_{N}(x) = {\alpha \over 2N(N-1) } \sum^{N-1}_{m=1} \{
[(10 K(m) - 2N^{2}] K(m) + \alpha  |x| [N^{2} - 4 K(m)] K(m)^{2} \}
w^{K(m)}
\end{equation}
where $K(m) = mN-m^{2}$.  Since $K(N-m) = K(m)$, the sum contains every term twice,
except for $m = N/2$ when $N$ is even.  The special value at $x = 0$ (which is the
probability of having two particles at the same place) is 
$C_{N}(0) = (N+1)/6$.  Figure 3 gives a graph of $C_{N}(x)/\alpha N^{2}$ for
various $N$.  In
the limit of large $N$, we may approximate $N^{2} \gg  K(m)$ in the prefactors and 
then
$K(m) \approx  Nm$, with the result
\begin {equation}
\label{Eq17}
C_{N}(x) \approx  {\alpha \over 4N } \{  N\alpha |x| {\cosh(N\alpha
x/2)\over \sinh^{3}(N\alpha |x|/2)} - {2\over
\sinh^{2}(N\alpha x/2)} \}
\end{equation}
The structure factor is given by the Fourier transform of 
$C_{N}(x)$:
\begin {equation}
\label{Eq:Eq18}
C_{N}(k) = {\alpha \over N(N-1) } \sum^{N-1}_{m=1} K(m)^{2} \{ {[10
K(m) - 2N^{2}] \over k^{2}+ K(m)^{2}} - {[N^{2} - 4 K(m)][K(m)^{2}-k^{2}]\over 
[k^{2}+ K(m)^{2}]^{2} } \}
\end{equation}
\section{Excited states}
In this section we will show that for any $N$ there is only one localized state 
with zero total momentum.  There are other states of negative energy, but 
these can all be interpreted as uncorrelated smaller molecules.  

For indistinguishable bosons we need only specify the wavefunction for the 
sector $x_{1} \le  x_{2} \le  \ldots   \le  x_{N}$, and in the interior of this 
region (where no two 
coordinates are equal), the Hamiltonian operator reduces to the kinetic 
energy.  The bound eigenstates of the kinetic energy have the form
\begin {equation}
\label{Eq19}
\Psi (\{x_{i}\};\{k_{i}\}) = \exp(\sum  k_{i}x_{i}) ,
\end{equation}
and the energy is determined to be $E = -\sum^{N}_{i=1} \hbar ^{2}k^{2}/2$m.  To be 
normalizable it is
necessary that $\sum^{N}_{i=1} k_{i} = 0$ (otherwise a uniform translation of all the
particles could lead to indefinite growth of $\Psi$ ).  The wavefunction for the
interacting system is a linear combination of degenerate functions of this
form.  At the sector boundaries 
(where $x_{j}=x_{j+1}$ for some j), the interaction gives
rise to a discontinuity in the derivative of the wavefunction, which
implies\cite{Lieb}
\begin {equation}
\label{Eq20}
(\partial /\partial x_{j+1} - \partial /\partial x_{j} ) \Psi (\{x_{i}\})_{|x_{j+1}}{ 
} _{=}{ } _{x_{j}}= -\alpha  \Psi (\{x_{i}\})_{|x_{j+1}}{ } _{=}{ } _{x_{j}}
\end{equation}
This must hold for arbitrary choices of the remaining $N-2$ variables; then in 
the sum of terms of the form (\ref{Eq19}) the only terms that are coupled by 
(\ref{Eq20}) are those for which the lists $\{k_{i}\}$ differ only at $k_{j}$ and 
$k_{j-1}$ -- 
but then the two sum constraints on the $\{k_{i}\}$ ensure that there are only two
such lists, which differ by having the values for $k_{j}$ and $k_{j-1}$ interchanged.
Thus the general form of the wavefunction would appear to be constructable
from a single list $\{k_{i}\}$ (which we will take to be ordered, so that $k_{i} > 
k_{j}$ for
all $i < j)$ in the form
\begin {equation}
\label{Eq21}
\Psi (\{x_{i}\}) = \sum^{}_{P} C_{P} \exp(\sum^{N}_{i=1} x_{i} k_{P(i)})  ,
\end{equation}
where the first sum is over all permutations of $N$ objects, and $P(i)$ is the 
$i^{th}$
member of the permuted list.  The boundary condition (\ref{Eq20}) implies
relations among the coefficients $C_{P}$
\begin {equation}
\label{Eq22}
(\alpha  - k_{P(j)} + k_{P(j+1)}) C_{P} = - (\alpha  - k_{Q(j)}+k_{Q(j+1)}) C_{Q} 
\equiv  -(\alpha  + k_{P(j)} - k_{P(j+1)})
C_{Q}
\end{equation}
(It should be noted that the permutations $P$ and $Q$ are related so that $P(j) =
Q(j+1)$ and $P(j+1)=Q(j))$.

The foregoing differs only slightly from the case of bosons with repulsive 
delta-potential interactions\cite{Lieb} -- the main difference is that the $k_{i}$ 
are
real instead of purely imaginary.  However, the remaining boundary conditions
are different: for repulsive interactions, the wavefunction is delocalized and
must be confined within a periodic box, while for attractive interactions, the
space can be infinite if the wavefunction itself is localized.

The relevant (and interesting) condition is that $\Psi (\{x_{i}\})$ should become
exponentially small if the extremal particle at $x_{N}$ is moved to large positive
$x$, or if the particle at $x_{1}$ is moved to large negative $x$.  
This requires that
$k_{P(1)}$ be positive and $k_{P(N)}$ be negative, for every term in (\ref{Eq21}) 
that
has nonzero coefficient.  The sum itself is over all permutations, which
certainly will generate terms in which the k's have the wrong sign; but the
coefficients will vanish (according to (\ref{Eq22})) for all permutations such
that in the list $\{k_{P(i)}\}$ it happens there are members $k_{P(i)}$ and 
$k_{P(j)}$ such
that $i < j$ and $k_{P(j)} = k_{P(i)} - \alpha $.  The ground state wavefunction is a 
simple
example of the working of this rule: the ordered list is of the form of a
sequence  $(k_{1},k_{1}-\alpha ,k_{1}-2\alpha ,k_{1}-3\alpha ,\ldots  )$ and every 
possible permutation of it gives
rise to a vanishing coefficient.

The ground state is not the only possibility: we could have a list which 
is made of several distinct sequences, so that in
one of its permutations it takes the form 
$(k_{1},k_{1}-\alpha ,\ldots  k_{a}-n_{a}\alpha ,k_{b},k_{b}-\alpha ,..k_{b}-
n_{b}\alpha ,k_{c},k_{c}-\alpha \ldots  )$.  This gives some freedom in
the choice for the starting elements $k_{p}$ and sublist lengths $n_{p}$; this 
generates
the wavefunction for an excited state of the molecule.  However, it does not
represent a localized state.  Suppose we simultaneously displace the
particles $\{x_{i}: i =1,2,..,n_{a}+1\}$ a large distance to negative values.   If 
$\sum^{n_{a}+1}_{i=1}
k_{i} < 0$, this term will become exponentially large, giving a nonnormalizable
wavefunction.  However, if $\sum^{n_{a}+1}_{i=1} k_{i} > 0$ there will be a different 
permutation
of the list in which this sublist is at the end, again giving a
nonnormalizable wavefunction when these particles are moved to large positive
values.  The remaining possibility is $\sum^{n_{a}+1}_{i=1} k_{i} = 0$ -- but now the 
wavefunction
remains finite if it is separated into two widely separated pieces, one of
which contains $n_{a}+1$ particles: this describes states in which the particles
have formed several separate and uncorrelated molecules.

In the foregoing argument we have assumed that the $\{k_{i}\}$ are real.  Allowing 
them to be complex does not change the situation much: the argument above 
implies that the k$_{i}$ having a real part must be organizable into sublists as 
above, all having the same imaginary part; each sublist represents a separated 
molecule, now with finite momentum.  From this point of view we can readily 
see that when two molecules collide, nothing happens -- they pass through each 
other leaving each undisturbed\cite{Yoon}.
 
{\bf Acknowledgement} 
 
This work was supported by the Thomas F. Jeffress and Kate Miller
Jeffress Memorial Trust.

\leftline{{\bf Figure Caption}}
\begin{figure}
\centerline {\epsfxsize=3.in\epsfbox{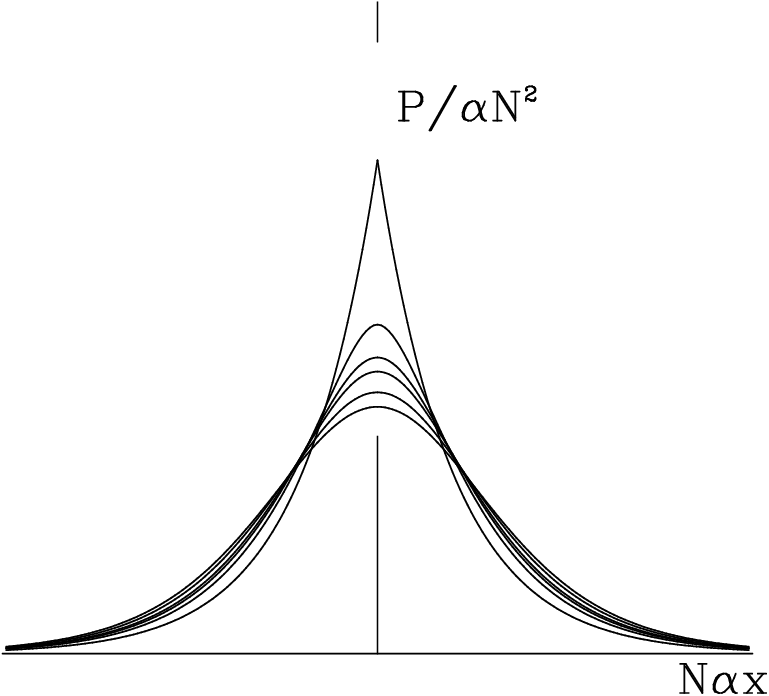}}
\vspace{3mm}
\caption{
Particle density distribution.  Here are displayed 
$ P_{N}(x)/ \alpha N^{2}$ as a function of
$N \alpha x $, 
for $N=2,3,4,5,10$, and $\infty $.  The cusp in
$P_{2}$ at y = 0
is replaced by discontinuities in the higher derivatives for other N.
}
\end{figure}
\begin{figure}
\centerline {\epsfxsize=3.in\epsfbox{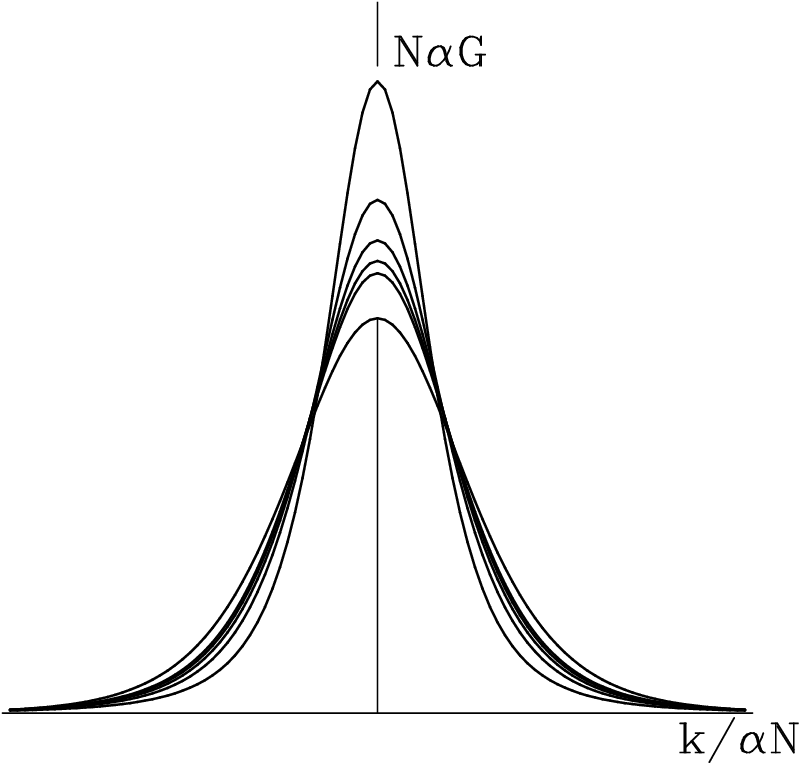}}
\vspace{3mm}
\caption{
The single-particle momentum distribution.  Here are displayed
$N \alpha G_{N}(k)$  as a function of $k / \alpha N$, for $N = 2, 3, 4,
5, 6,$ and $50$.
}
\begin{figure}
\centerline {\epsfxsize=3.in\epsfbox{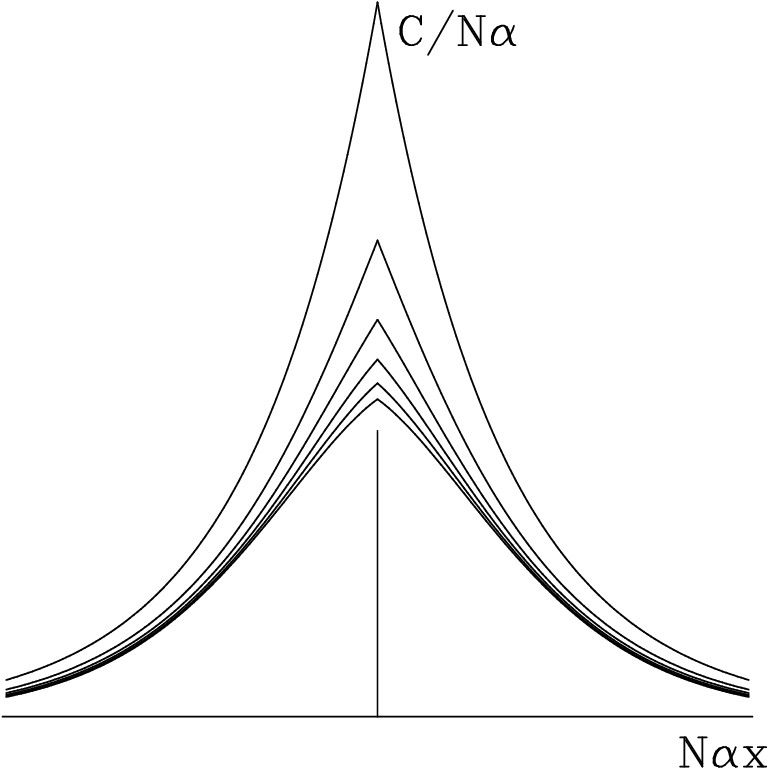}}
\vspace{3mm}
\caption{
The two-point correlation function.  Here are displayed 
$C_{N}(x)/\alpha N^{2}$ as a function of $N \alpha x$,
for $N = 2, 3, 4, 5, $ and $6$.}
\end{figure}
\end{figure}

\begin{references}
\bibitem{note} It's a little surprising that 
there is any singularity, since placing one 
particle at the center of mass would not seem to be significantly different 
from placing it nearby.  However, for this special position the configurations 
with differing numbers of particles to right and left of the origin (for $N$ 
even, {\sl all }configurations have this property) are equivalent and coupled by the 
possibility of moving another particle from one side to the other.
\end{references}
\end{document}